\documentclass[12pt]{article}
\usepackage[dvips]{graphicx}
\usepackage{color}
\usepackage{amsfonts,amssymb}
\begin{document}

\vspace*{1cm}
\begin{center}
{\Large \bf The SELEX Measurements\\[1ex]
in the Unified Picture for Hadron Spectra}\\

\vspace{2mm}

{\large A.A. Arkhipov\footnote{e-mail: arkhipov@mx.ihep.su}\\
{\it State Research Center ``Institute for High Energy Physics" \\
 142281 Protvino, Moscow Region, Russia}}\\
\end{center}

\vspace{2mm}
\begin{abstract}
We give an analysis of the experimental material presented by the
SELEX Collaboration to find the true place for the SELEX state
$D_{sJ}^+$(2632) in the unified picture for hadron spectra developed
early. It is found that the SELEX measurements are excellently
incorporated in the unified picture for hadron spectra. Our analysis
shows that the measured values for the masses of the SELEX state
exactly coincide with the calculated masses of the states living in
the corresponding KK towers. We also found quite a natural but rather
model dependent explanation of the decay pattern for the SELEX state
being dominated by the $D_s^+\eta$ decay mode.
\end{abstract}

\section{Introduction}

Exciting experimental measurements in hadron spectroscopy of last two
years clearly open a new page in hadron physics which is certainly a
starting page of a new era in particle physics. Indeed, there were
discovered many new narrow hadronic states with unexpected
properties. First of all, a series of new mesons have been discovered
whose properties are in a strong disagreement with the predictions of
conventional QCD-inspired quark potential models. New narrow meson
$D_{sJ}^{*+}(2317)$, decaying into $D_s^+\pi^0$ has been observed by
BABAR Collaboration \cite{1} in the first. This observation was soon
confirmed by CLEO Collaboration \cite{2}, which have also established
the existence of a new narrow state with a mass near 2.46 GeV in its
decay to $D_s^{*+}\pi^0$. Belle Collaboration \cite{3} reported the
first observation of the $D_{sJ}(2317)$ and $D_{sJ}(2457)$ in $B$
decays: $B \rightarrow \bar DD_{sJ}(2317)$ and $B \rightarrow \bar
DD_{sJ}(2457)$ with a subsequent $D_{sJ}(2317)$ decay to $D_s\pi^0$
and $D_{sJ}(2457)$ decay to $D_s^*\pi^0$ and $D_s\gamma$ final
states. Both CLEO and Belle observations of $D_{sJ}(2457)$ have been
confirmed by BABAR \cite{4}. Moreover, Belle Collaboration has
reported \cite{5,6} the discovery of very narrow $X(3872)$-meson
state ($\Gamma_{X(3872)}^{tot}<2.3 MeV$) in the $J/\psi\pi^+\pi^-$
invariant mass distribution in the $B$ decay $B^{\pm} \rightarrow
K^{\pm}J/\psi\pi^+\pi^-$. This observation of Belle Collaboration was
soon confirmed by CDF at Fermilab \cite{7}. The mass measurement
presented by CDF 3871.4 $\pm$ 0.7 $\pm$ 0.4 MeV is in agreement with
the result of Belle. It should be noted, in particular, that the mass
2317 MeV is approximately 41 MeV below the $DK$ threshold but the
mass 3872 MeV is very near the $D^0{\bar D}^{*0}$ threshold, while
the $D^+D^{*-}$ channel with approximately 8 MeV higher threshold
mass is forbidden for $X(3872)$ decay by phase space. What is
remarkable here that all new narrow states have been observed at the
masses which are surprisingly far from the predictions of
conventional quark potential models. It is still more non-trivial
that all new observed states are very narrow, their total widths
being about a few MeV. The small widths were found to be in
contradiction with quark model expectations.

Quite recently the SELEX Collaboration \cite{8,9} has reported the
first observation of a charm-strange meson $D_{sJ}^+$(2632) in the
charm hadro-production experiment E781 at Fermilab. This state was
seen in two decay modes, $D_s^+\eta$ and $D^0K^+$. In the $D_s^+\eta$
decay mode an excess of 49.3 events with a significance of 7.2
$\sigma$ at a mass of 2635.9 $\pm$ 2.9 MeV has been observed. There
is a corresponding peak of 14 events with a significance of 5.3
$\sigma$ at a mass of 2631.5 $\pm$ 1.9 MeV in the decay mode
$D^0K^+$. Upper limit for the decay width of this state, consistent
with a width due only to resolution in the $D^0K^+$ decay mode, is
$<$ 17 MeV at 90$\%$ confidence level. The relative branching ratio
$\Gamma(D^0K^+)/\Gamma(D_s^+\eta)$ is 0.16 $\pm$ 0.06. The mechanism
which keeps this state narrow is unclear. Its decay pattern being
dominated by the $D_s^+\eta$ decay mode is also unusual to be
explained in the conventional quark models. From the standard quark
model point of view the $D_{sJ}^+$(2632) meson which is 660 MeV
heavier than the ground $D_s$ state and about 100 MeV below the quark
models predictions of 2.73-2.81 GeV for the 2${}^3S_1$ $c\bar s$
state together with the $D_{sJ}(2317)$ and $D_{sJ}(2457)$ states
mentioned above seem constitute quite a new group of the
charm-strange family rather distinguished from the ordinary
$D_{s}$(1968) and $D_{s}^*$(2112) mesons by unusual features in the
masses, narrow widths and decay pattern. It is quite unclear how
these newly discovered mesons could be understood as the states of a
simple $c\bar s$ quark system. In fact, a description of new,
recently discovered meson states with unexpectedly low masses, narrow
total decay widths and unusual decay pattern is extremely problematic
in the framework of phenomenology based on conventional quark models.
Why these meson states are less massive and so narrow than earlier
predicted in quark models is an open question so far. Certainly, all
of that raise a challenge to the theory.\footnote{A comprehensive
analysis of the different options for the SELEX state
$D_{sJ}^+$(2632) in quark models  has been made in \cite{10}.} In
other words, this means either considerable modifications in the
conventional quark models have to be introduced to reconsider the
systematics of charmed and charmed-strange mesons spectroscopy and
strong decays or that completely new approaches should be applied in
hadron spectroscopy.

Recent discoveries of very narrow, manifestly exotic baryons lead us
to the same conclusion. Here are the first observations of the
$\Theta^+$ (Q=1, S=1) states with the simplest quark assignment
($uudd\bar s$) decaying into $nK^+$ and $pK_S^0$ \cite{11} and the
discovery of another exotic baryon $\Xi^{--}$ (Q=--2, S=--2), now
denoted as $\Phi^{--}$ by PDG, with the quark assignment ($ddss\bar
u$) decaying into $\Xi^-\pi^-$ at the mass $M=1862$ MeV with the
width $\Gamma < 18$ MeV \cite{12}, but see also \cite{13}.

In Ref. \cite{14}, where some of our previous studies were partially
summarized, it has been claimed that existence of the extra
dimensions in the spirit of Kaluza and Klein together with some novel
dynamical ideas may provide new conceptual issues for the global
solution of the spectral problem in hadron physics to build up a
unified picture for hadron spectra. We have shown \cite{15}, in
particular, that all charmed and charmed-strange mesons, including
the recently observed new states, have excellently been incorporated
in the systematics provided by the created unified picture for hadron
spectra. A thorough analysis \cite{16} of many different experiments
reported the observation of a new $\Theta$ baryon taken together
allowed us to claim that many different $\Theta$ states have been
discovered and all of them were excellently incorporated in the
unified picture for hadron spectra developed early \cite{14}. In this
article we give the similar analysis of the experimental material
presented by SELEX Collaboration to find the true place for the SELEX
state $D_{sJ}^+$(2632) in the unified picture for hadron spectra.

\section{Understanding the SELEX state $D_{sJ}^+$(2632) \\
in the unified picture for hadron spectra}

The SELEX experiment \cite{8} have used the Fermilab charged hyperon
beam at 600 GeV to produce charmed particles in a set of thin foil
targets of Cu or diamond.  The negative beam composition was
approximately half $\Sigma^-$ and half $\pi^-$. The most important
features of the experiment were the high-precision, highly redundant,
vertex detector that provided an average proper time resolution of 20
fs for charm decays, a 10 m long Ring-Imaging Cerenkov (RICH)
detector that separated $\pi$ from K up to 165 GeV, and a
high-resolution tracking system that had momentum resolution of
$\sigma_{p}/p<1\%$ for a 150 GeV track.  Photons have been detected
in 3 lead glass photon detectors, one following each spectrometer
magnet. The photon angular coverage in the center of mass of a
production collision exceeded 2$\pi$.  For the presented analysis the
photon energy threshold was 2 GeV.

The experiment recorded data from $15.2 \times 10^{9}$ inelastic
interactions and wrote $10^{9}$ events to tape using both positive
and negative beams. The data set had 65\% of the events induced by
the $\Sigma^{-}$ beam with the balance split roughly equally between
$\pi^{-}$ and protons.  Previous SELEX $D_s$ studies have shown that
most of the signal came from the $\Sigma^-$ beam, therefore it was
restricted in the analysis within the $9.9 \times 10^{9}$
$\Sigma^{-}$-induced interactions. The details of the procedure how
charm candidates have been selected and many other details can be
found in original paper \cite{8} and references therein.

The results of the search for the $D_s\eta$ decay mode were presented
in the $M(KK\pi^\pm\eta) - M(KK\pi^\pm)$ mass difference distribution
shown in Fig.~1 extracted from the original paper. In this plot the
$\eta$ mass was fixed at the PDG value. A clear peak is seen for a
mass difference of 667.4 $\pm$ 2.9 MeV. There is an excess of 43.9
events over an expected background of 51.7 events with a significance
of 7.2 $\sigma$ at a mass of 2635.9 $\pm$ 2.9 MeV. The signal did not
change with  variations of $\pm$ 2$\%$ in the photon energy scale.

The results of the search for the $D^0K^+$ decay mode were shown in
the $M(K^-\pi^+K^+) - M(K^-\pi^+)$ mass difference distribution
depicted in Fig.~2 extracted from the original paper too. The known
$D_{sJ}$(2573) state is clearly seen in this Figure, but there is
another peak above the $D_{sJ}$(2573). Both of the peaks were fitted
with a Breit-Wigner convolved with a fixed width Gaussian using a
constant background term. The Gaussian resolution was set to the
simulation value of 4.9 MeV. The mass difference and width of the
$D_{sJ}$(2573) obtained by the fit agreed well with the PDG values.
The fitted mass difference of the second Breit-Wigner was obtained of
767.0 $\pm$ 1.9 MeV, leading to a mass for the new peak of 2631.5
$\pm$ 1.9 MeV.  For the Breit-Wigner width it was found that it was
$<$ 17 MeV at 90$\%$ confidence level. This signal has a significance
of 5.3 $\sigma$. It was also pointed out that the mass difference
between this signal and the one seen in the $D_s\eta$ mode is 3.2
$\pm$ 3.5 MeV, statistically consistent with being the same mass. For
the relative branching ratio $\Gamma(D^0K^+)/\Gamma(D_s^+\eta)$ the
value 0.16 $\pm$ 0.06 has been reported.

In conclusion the SELEX reported the measured peaks as the first
observation of yet another narrow, high-mass $D_s$ state decaying
strongly to a ground state charm plus a pseudoscalar meson.  The
mechanism which keeps this state narrow is unclear. The branching
ratios for this state are also unusual. The $D_s\eta$ decay rate
dominates the $D^0K^+$ rate by a factor of $\sim$~6 despite having
half the phase space. To place this new state in the spectroscopy of
the charm-strange meson system will require careful study from a
number of experiments in the future.

Recently we suggested quite a new scheme of systematics for hadron
states. The fundamental Kaluza-Klein hypothesis on existence of the
extra dimensions with a compact internal extra space is a base of our
approach to hadron spectroscopy. The observed hadron states occupy
the storeys and live in the corresponding KK towers built in
according to the established general physical law. Herewith the size
of the internal compact extra space determines the global
characteristics of the hadron spectra while the masses of the decay
products are the fundamental parameters of the compound systems which
appear as the elements of the global structure. Our approach to
hadron spectroscopy has been verified with a large amount of
experimental data on hadron states and received an excellent
agreement; see \cite{14,16} and references therein for the details.
What is remarkable that all new hadron states experimentally
discovered last two years have been observed just at the masses
predicted in our approach, and those states appeared to be narrow as
predicted too. Here we apply our approach to analyse the experimental
material presented by SELEX Collaboration.

First of all, it should be emphasized that the Kaluza-Klein tower of
KK-excitations for the $DK$ system has already been built in Ref.
\cite{15}. The built Kaluza-Klein tower extracted from Ref. \cite{15}
is shown in Table 1. Clearly, the SELEX state shown by bold-face
number in Table 1 just occupies the 12th storey in the Kaluza-Klein
tower. Some known experimental information has been presented in this
Table as well.

We have also built the Kaluza-Klein tower of KK-excitations for the
$D_s\eta$ system by the formula

\begin{equation}\label{Dseta}
M_n^{D_s\eta} = \sqrt{m_{D_s}^2+\frac{n^2}{R^2}} +
\sqrt{m_{\eta}^2+\frac{n^2}{R^2}},\quad (n=1,2,3,...),
\end{equation}
where $R$ is the same fundamental scale established before; see
\cite{14} and references therein for the details. The such built
Kaluza-Klein tower is shown in Table 2. As seen the SELEX measured
state just lives in 8th storey within this Kaluza-Klein tower.

Here we would like to compare the theoretically calculated spectra
with the experimental material presented by SELEX Collaboration.

The $D^0K^+$ invariant mass spectrum measured by SELEX is shown in
Fig. 3 extracted from Ref. 9.\footnote{I thank A. Evdokimov for
drawing my attention to Ref. 9 and sending me the pictures from his
talk.} We have plotted in Fig. 3 the spectral lines corresponding to
KK excitations in the $D^0K^+$ system taken from Table 1. As it
should be expected the spectral line corresponding to the
$M_{12}^{D^0K^+}$(2630.5)-storey exactly coincided with a clear peak
on the histogram. Here is also clear seen a strong correlation of the
spectral lines with the other experimentally observed peaks, and we
already have early found out that strong correlation more than once
in our previous studies \cite{16}. At least, it should be emphasized
an evidence for the two other states corresponding to the spectral
lines $M_{10}^{D^0K^+}$(2554.5) and $M_{11}^{D^0K^+}$(2591.4) with
clear peaks on the histogram in Fig. 3. Obviously, a further, much
more careful experimental studies with a higher statistics and better
mass resolution are very desired.

Figure 4 extracted from Ref. 9 too shows the $D_s^+\eta$ invariant
mass spectrum measured by SELEX Collaboration as well. We have also
plotted in Fig. 4 the spectral lines corresponding to KK excitations
in the $D_s^+\eta$ system taken from Table 2. Again we found a
remarkable correlation of the spectral lines with the peaks on the
histogram. The reported state by SELEX just corresponds to the
spectral line from $M_{8}^{D_s^+\eta}$(2636.8)-storey in KK tower for
the $D_s^+\eta$ system given by Table 2. Here a further, careful
experimental studies with a higher statistics and better mass
resolution are also utterly important.

\section{What could we say about the decay widths?}

Our conservative estimate for the widths of KK excitations looks like
\cite{17}
\begin{equation}\label{width}
\Gamma_n \sim \frac{\alpha}{2}\cdot\frac{n}{R}\sim 0.4\cdot n\,
\mbox{MeV},
\end{equation}
where $n$ is the number of KK excitation, and $\alpha \sim 0.02$,
$R^{-1}=41.48\,\mbox{MeV}$ have been taken from our previous studies
\cite{17,14}. This gives $\Gamma_{12}(D^0K^+)\sim 4.8$ MeV and
$\Gamma_{8}(D_s^+\eta)\sim 3.2$ MeV. Thus, in that case, the ratio
\begin{equation}\label{ratio}
\frac{\Gamma_{12}(D^0K^+)}{\Gamma_8(D_s^+\eta)}=1.5
\end{equation}
is approximately 10 times larger than the experimental value.
However, it's clear, that account of some additional, model dependent
dynamical assumptions may change that estimate. For this goal let us
consider the simplest multidimensional model with an effective
interaction
\begin{equation}\label{action}
S_I = \int_{{\cal M}_{(4+d)}} d^{4+d}z\left[G_{(4+d)}^{D^0K}
\Phi_{D^0}^2(z)\Phi_{K}(z)\Phi_{K}^*(z) + G_{(4+d)}^{D_s\eta}
\Phi_{\eta}^2(z)\Phi_{D_s}(z)\Phi_{D_s}^*(z)\right],
\end{equation}
where ${\cal M}_{(4+d)} = M_4 \times {\cal K}_d$, $M_4$ is the
pseudo-Euclidean Minkowski space-time, ${\cal K}_d$ is a compact
internal $d$-dimensional extra space with the characteristic size
$R$, $z^M = \{ x^{\mu}, y^{m}\}$, ($M=0,1,\ldots,3+d,\, \mu =
0,1,2,3,\, m=1,2, \ldots, d$) are local coordinates on ${\cal
M}_{(4+d)}$ so that $x^{\mu}\in M_4$, $y^{m}\in {\cal K}_d$,
$\Phi_{P}(z)$, ($P=D^0,K,D_s,\eta$), are local multidimensional
fields corresponding to the decay products. We choose a
$d$-dimen\-sional torus ${\cal T}^{d}$ with equal radii $R$ as an
especially simple example of the compact internal space of extra
dimensions ${\cal K}_d$. The eigenfunctions and eigenvalues of the
Laplace operator on the internal space ${\cal K}_d$ in this special
case have quite a simple analytical form
\begin{equation}
\Delta_{{\cal K}_{d}} Y_{n}(y) = -\frac{\lambda_{n}}{R^{2}} Y_{n}(y),
\quad Y_n(y) = \frac{1}{\sqrt{V_d}} \exp \left(i \sum_{m=1}^{d}
n_{m}y^{m}/R \right), \label{Yn}
\end{equation}
\[
\lambda_n = |n|^2,\quad |n|^2= n_1^2 + n_2^2 + \ldots n_d^2, \quad
n=(n_1,n_2, \ldots, n_d),\quad -\infty \leq n_m \leq \infty,
\]
where $n_m$ are integer numbers, $V_d = (2\pi R)^d$ is the volume of
the torus.

We write a harmonic expansion for the multidimensional fields
$\Phi_P(z)$ to reduce the multidimensional theory to the effective
four-dimensional one
\begin{equation}
\Phi_P(z) = \Phi_P(x,y) = \sum_{n} \phi_P^{(n)}(x) Y_{n}(y).
\label{H}
\end{equation}
The coefficients $\phi_P^{(n)}(x)$ of harmonic expansion (\ref{H})
are called Kaluza-Klein (KK) excitations or KK modes, and they
include the zero-mode $\phi_P^{(0)}(x)$, corresponding to $n=0$ and
the eigenvalue $\lambda_{0} = 0$. Substitution of the KK mode
expansion into action (\ref{action}) and integration over the
internal space ${\cal K}_{d}$ gives
\begin{equation}
S_I = \int_{M_4} d^4x\left[g_{D^0K}
(\phi_{D^0}^{(0)}(x))^2\phi_{K}^{(0)}(x)\phi_{K}^{(0)*}(x) +
g_{D_s\eta}
(\phi_{\eta}^{(0)}(x))^2\phi_{D_s}^{(0)}(x)\phi_{D_s}^{(0)*}(x)\right.
+  \label{S}
\end{equation}
\[
+\left. g_{D^0K} \phi_{D^0}^{(0)}(x)\phi_{K}^{(0)}(x) \sum_{n\neq 0}
\phi_{D^0}^{(n)}(x) \phi_{K}^{(n)*}(x) + g_{D_s\eta}
\phi_{\eta}^{(0)}(x)\phi_{D_s}^{(0)}(x) \sum_{n\neq 0}
\phi_{\eta}^{(n)}(x) \phi_{D_s}^{(n)*}(x) \right] + \ldots.
\]
The coupling constants $g$ of the four-dimensional theory are related
to the coupling constants $G_{(4+d)}$ of the initial multidimensional
theory by the equation
\begin{equation}
g_{D^0K} = \frac{G_{(4+d)}^{D^0K}}{V_d}, \quad g_{D_s\eta} =
\frac{G_{(4+d)}^{D_s\eta}}{V_d},\label{g}
\end{equation}
where $V_d$ is the volume of the compact internal extra space ${\cal
K}_d$. The fundamental coupling constants $G_{(4+d)}$ have dimension
$[\mbox{mass}]^{-d}$. So, the four-dimensional coupling constants $g$
are dimensionless, as it should be.

As we have established before \cite{17}
\begin{equation}\label{size}
\frac{1}{R}=41.481 \mbox{MeV}\quad \mbox{or}\quad
R=24.1\,GeV^{-1}=4.75\,10^{-13}\mbox{cm}\,.
\end{equation}
If we relate the strong interaction scale with the pion mass
\begin{equation}\label{G}
G_{(4+d)}\sim\frac{10}{[m_\pi]^d},
\end{equation}
then
\begin{equation}\label{simg}
g\sim\frac{10}{(2\pi m_\pi R)^d},
\end{equation}
and
\[
g(d=1)\sim 0.5.
\]
On the other hand
\begin{equation}\label{geff}
g_{eff}=g_{\pi NN}\exp(-m_{\pi}R)\sim 0.5,\ \ \  (g^2_{\pi
NN}/4\pi=14.6).
\end{equation}
So, $R$ has a clear physical meaning: size (\ref{size}) just
corresponds to the scale of distances where strong Yukawa forces in
strength come close to electromagnetic ones \cite{14,17}. Physically
this means that KK modes of strong interacting particles interact
weakly. This fact allows us to apply the four-dimensional
perturbation theory in the calculations and to use a quantum
mechanical Schr\"{o}dinger wave function to describe the
configuration of a compound system with a non-zero KK modes as
constituents. By this way in the lowest order over coupling constant
in the above model we can easily calculate the decay widths. As a
result one obtains
\begin{equation}\label{Gamma1}
\Gamma(D_{sJ}\rightarrow D^0K^+) = C_0
\frac{\alpha_{D^0K}}{M_{D_{sJ}}^2}\cdot|\,\psi_{Sch}^{D_n^0K_n^+}(0)|^2,\quad
\alpha_{D^0K}=\frac{g_{D^0K}^2}{4\pi},
\end{equation}
and
\begin{equation}\label{Gamma2}
\Gamma(D_{sJ}\rightarrow D_s^+\eta) = C_0
\frac{\alpha_{D_s\eta}}{M_{D_{sJ}}^2}\cdot|\,\psi_{Sch}^{D_{s,n}^+\eta_n}(0)|^2,\quad
\alpha_{D_s\eta}=\frac{g_{D_s\eta}^2}{4\pi},
\end{equation}
where $C_0$ is known, model dependent constant.

Now, let us consider a Coulomb-like wave function of the ground state
as a Schr\"{o}dinger wave function describing the configuration of
the non-zero KK modes in a compound system
\begin{equation}\label{config1}
\psi_{C}^{D_n^0K_n^+}(\vec x) =
\frac{1}{\sqrt{\pi}}\left(\frac{1}{2a_n}\right)^{3/2}\exp(-|\vec
x|/2a_n) =
\frac{1}{\sqrt{\pi}}\left(\frac{\alpha_{D^0K}\mu_n}{2}\right)^{3/2}\exp(-\alpha_{D^0K}\mu_n|\vec
x|/2),
\end{equation}
\[
a_n = \frac{1}{\alpha_{D^0K}\mu_n},\quad \mu_n =
\frac{m_n^{D^0}m_n^{K^+}}{m_n^{D^0}+m_n^{K^+}},\quad
m_n^{D^0}=\sqrt{m_{D^0}^2+\frac{n^2}{R^2}},\quad
m_n^{K^+}=\sqrt{m_{K^+}^2+\frac{n^2}{R^2}},
\]
and the similar wave function for the $D_{s,n}^+\eta_n$ configuration
\begin{equation}\label{config2}
\psi_{C}^{D_{s,n}^+\eta_n}(\vec x) =
\frac{1}{\sqrt{\pi}}\left(\frac{1}{2{\tilde
a}_n}\right)^{3/2}\exp(-|\vec x|/2{\tilde a}_n) =
\frac{1}{\sqrt{\pi}}\left(\frac{\alpha_{D_s\eta}\tilde\mu_n}{2}\right)^{3/2}\exp(-\alpha_{D_s\eta}\tilde\mu_n|\vec
x|/2),
\end{equation}
\[
{\tilde a}_n = \frac{1}{\alpha_{D_s\eta}\tilde\mu_n},\quad
\tilde\mu_n =
\frac{m_n^{D_s^+}m_n^{\eta}}{m_n^{D_s^+}+m_n^{\eta}},\quad
m_n^{D_s^+}=\sqrt{m_{D_s^+}^2+\frac{n^2}{R^2}},\quad
m_n^{\eta}=\sqrt{m_{\eta}^2+\frac{n^2}{R^2}}.
\]
From Eqs. (\ref{Gamma1}-\ref{config2}) it follows for the SELEX state
\begin{equation}\label{ratio}
\frac{\Gamma(D_{sJ}\rightarrow D^0K^+)}{\Gamma(D_{sJ}\rightarrow
D_s^+\eta)} = \left(\frac{\alpha_{D^0K}}{\alpha_{D_s\eta}}\right)^4
\cdot\left(\frac{\mu_{12}}{\tilde\mu_8}\right)^3 = 1.194\cdot
\left(\frac{\alpha_{D^0K}}{\alpha_{D_s\eta}}\right)^4.
\end{equation}
The SELEX measured value for the ratio $\Gamma(D_{sJ}\rightarrow
D^0K^+)/\Gamma(D_{sJ}\rightarrow D_s^+\eta)$ is obviously achieved if
we take
\begin{equation}\label{rationum}
\frac{\alpha_{D^0K}}{\alpha_{D_s\eta}} = 0.6\quad \mbox{or}\quad
\frac{g_{D^0K}}{g_{D_s\eta}} = 0.77.
\end{equation}

In conclusion of this section, we would like to emphasize that the
unusual decay pattern observed by SELEX Collaboration, which is the
most problematic in conventional quark models, can easily be
explained in the unified picture for hadron spectra. No doubt, it
would be very interesting to construct such, more general, model
which the ratio (\ref{rationum}) appeared to be as an intrinsic
feature for.

\section{Summary and Discussion}

In this article we have presented an analysis of the experimental
material recently reported by the SELEX Collaboration. It is shown
that the SELEX measurements are excellently incorporated in the
unified picture for hadron spectra developed early. The main
advantage of our approach to hadron spectroscopy is that all
calculated numbers for the masses of hadron states do not depend on a
special dynamical model but follow from fundamental hypothesis on
existence of the extra dimensions with a compact internal extra
space. Our analysis shows that the measured values for the masses of
the SELEX state exactly coincide with the calculated masses of the
states living in the corresponding KK towers. This is really a
wonderful fact that the SELEX measurements even discriminate the
masses of the discovered state for the different configurations which
our systematics of hadron spectroscopy ascribe to, even though the
existing statistics is not sufficiently well to allow a more reliable
statement concerning that discrimination.

In our approach we also find quite a natural but rather model
dependent explanation of decay pattern for the SELEX state being
dominated by the $D_s^+\eta$ decay mode which is known as a great
puzzle in the framework of conventional quark model phenomenology.
Unfortunately, there is no other experiment with a confirmation of
the SELEX measurements, but we hope that new experiments with better
statistics and with higher mass resolution will appear in the near
future to confirm these exciting measurements. In this respect we
have paid attention to note made by SELEX Collaboration in original
paper concerning a one-bin excess in the $D^0K^+$ invariant mass
spectrum at 2636 MeV observed by the CLEO Collaboration \cite{18} ten
years ago. That's why we have also performed the spectral analysis of
the CLEO measurements. The result of our analysis is presented in
Fig.~5. As is seen from Fig.~5, there is really irregularity in the
$D^0K^+$ invariant mass spectrum around the $M_{12}^{D^0K^+}$(2630.5)
spectral line. The peak at 2573 MeV in the $D^0K^+$ invariant mass
spectrum reported by the CLEO Collaboration is near the
$M_{11}^{D^0K^+}$(2591) spectral line. The other irregularities in
the $D^0K^+$ invariant mass spectrum observed by the CLEO
Collaboration are strongly correlated with the calculated spectral
lines. The most striking manifestation of such correlation is exact
coincidence of the $M_{4}^{D^0K^+}$(2392) spectral line with the
observed very narrow peak at 2392 MeV. Here, it should be stressed
that the similar coincidence has been observed by the SELEX
Collaboration as well, see Fig.~3. Figure 3 also shows that the broad
peak at 2573 MeV in the $D^0K^+$ invariant mass spectrum reported by
the CLEO Collaboration, in fact, contains two peaks corresponding to
the $M_{10}^{D^0K^+}$(2554) and $M_{11}^{D^0K^+}$(2591) spectral
lines. The further study of this signal region with improved
statistics and better resolution is also necessary to confirm the
double peak structure.

In summary, newly performed experimental studies by the SELEX
Collaboration provide new, additional and excellent confirmation of
our theoretical conception. Of course, this fact is quite fascinating
and encouraging for us.


\newpage
\vspace*{2cm}
\begin{center}
Table 1: Kaluza-Klein tower of KK excitations for $DK$ system
\cite{15} \\and experimental data.

\vspace{5mm}
\begin{tabular}{|c|c|c|c|}\hline
 n & $ M_n^{D^\pm K^0}$\,MeV & $ M_n^{D^0K^\pm}$\,MeV & $
 M_{exp}^{DK}$\,MeV   \\
 \hline
1  & 2369.26 & 2359.98 &  \\
2  & 2375.78 & 2366.54 &  \\
3  & 2386.53 & 2377.37 &  \\
4  & 2401.35 & 2392.28 & $D_{sJ}$(2392) \\
5  & 2420.03 & 2411.08 &  \\
6  & 2442.33 & 2433.51 &  \\
7  & 2468.00 & 2459.32 & $D_{sJ}$(2463)? \\
8  & 2496.79 & 2488.25 &  \\
9  & 2528.45 & 2520.06 &  \\
10 & 2562.75 & 2554.51 &  \\
11 & 2599.47 & 2591.38 & $D_{sJ}^\pm$(2573) \\
12 & 2638.42 & 2630.48 & $\mathbf{2631.5 \pm 1.9}$ \\
13 & 2679.43 & 2671.64 &  \\
14 & 2722.34 & 2714.68 &  \\
15 & 2767.00 & 2759.48 &  \\
16 & 2813.29 & 2805.90 &  \\
17 & 2861.09 & 2853.84 &  \\
18 & 2910.32 & 2903.19 &  \\
19 & 2960.87 & 2953.86 &  \\
20 & 3012.67 & 3005.78 &  \\
21 & 3065.64 & 3058.86 &  \\
22 & 3119.73 & 3113.06 &  \\
23 & 3174.85 & 3168.29 &  \\
24 & 3230.98 & 3224.52 &  \\
25 & 3288.04 & 3281.69 &  \\
26 & 3346.00 & 3339.75 &  \\
27 & 3404.82 & 3398.65 &  \\
28 & 3464.44 & 3458.37 &  \\
29 & 3524.84 & 3518.87 &  \\
30 & 3585.99 & 3580.10 &  \\ \hline
\end{tabular}
\end{center}

\newpage
\vspace*{2cm}
\begin{center}
Table 1: Kaluza-Klein tower of KK excitations for $D_s\eta$ system
and the SELEX state.

\vspace{5mm}
\begin{tabular}{|c|c|c|}\hline
 n & $ M_n^{D_s^{\pm}\eta}$ MeV & $ M_{exp}^{D_s^{\pm}\eta}$ MeV  \\
 \hline
1  & 2518.31 &   \\
2  & 2524.30 &   \\
3  & 2534.20 &   \\
4  & 2547.88 &   \\
5  & 2565.18 &   \\
6  & 2585.90 &   \\
7  & 2609.85 &   \\
8  & 2636.82 & $\mathbf{2635.9\pm 2.9}$  \\
9  & 2666.58 &   \\
10 & 2698.96 &   \\
11 & 2733.74 &   \\
12 & 2770.75 &   \\
13 & 2809.84 &   \\
14 & 2850.85 &   \\
15 & 2893.64 &   \\
16 & 2938.10 &   \\
17 & 2984.11 &   \\
18 & 3031.58 &   \\
19 & 3080.41 &   \\
20 & 3130.53 &   \\
21 & 3181.86 &   \\
22 & 3234.32 &   \\
23 & 3287.87 &   \\
24 & 3342.44 &   \\
25 & 3397.99 &   \\
26 & 3454.46 &   \\
27 & 3511.81 &   \\
28 & 3570.00 &   \\
29 & 3629.00 &   \\
30 & 3688.76 &   \\ \hline
\end{tabular}

\end{center}

\newpage

\begin{figure}[htb]
\begin{center}
\includegraphics[width=\textwidth]{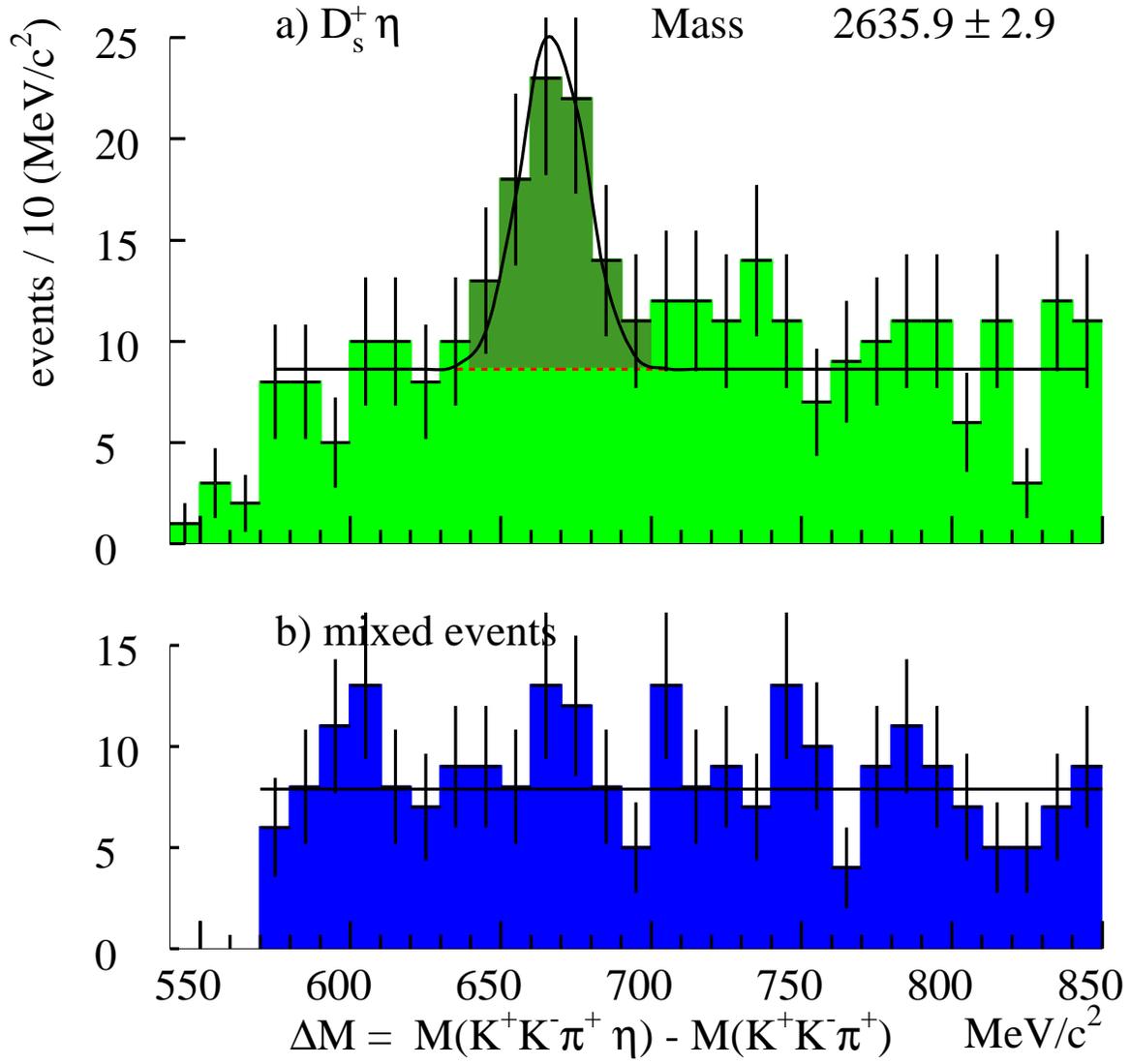}\label{fig1}
\end{center}
\caption{(a) $M(KK\pi^\pm\eta) - M(KK\pi^\pm)$ mass difference
distribution. Charged conjugates are included. The shaded region is
the event excess used in the estimation of signal significance.
Results for the fit shown see in original paper \cite{8}. (b) Mass
difference distribution for mixed events as described in the original
paper.}
\end{figure}

\newpage

\begin{figure}[htb]
\begin{center}
\includegraphics[width=\textwidth]{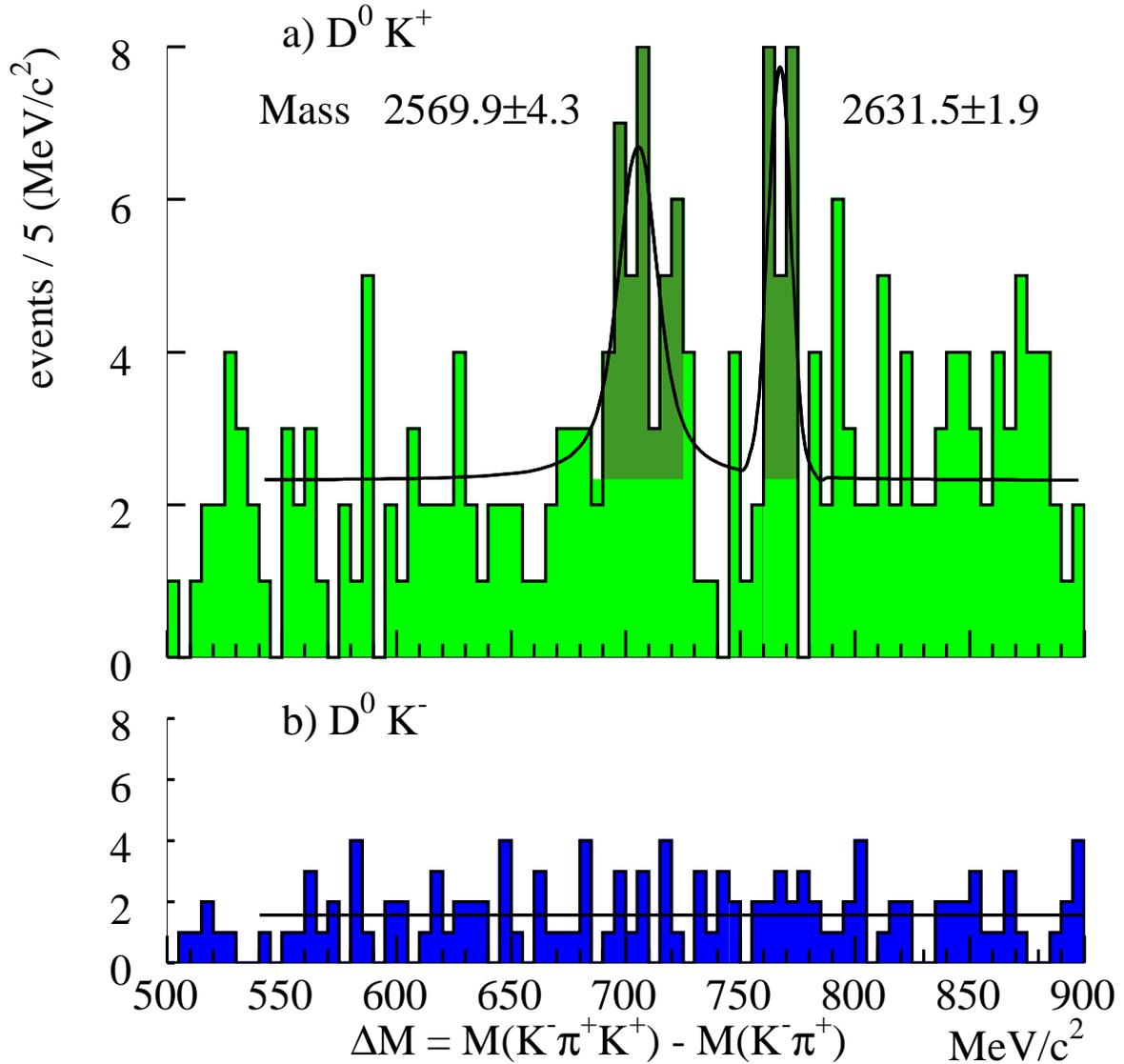}\label{fig2}
\end{center}
\caption{(a) $M(K^-\pi^+K^+) - M(K^-\pi^+)$ mass difference
distribution. Charged conjugates are included. The shaded regions are
the event excesses used in the estimation of signal significances.
Results for the fit shown see in original paper \cite{8}. (b) Wrong
sign background $\rm{D^0 \ K^{-}}$ events, as described in the
original paper.}
\end{figure}

\newpage

\begin{figure} [htb]
\begin{center}
\includegraphics[width=\textwidth]{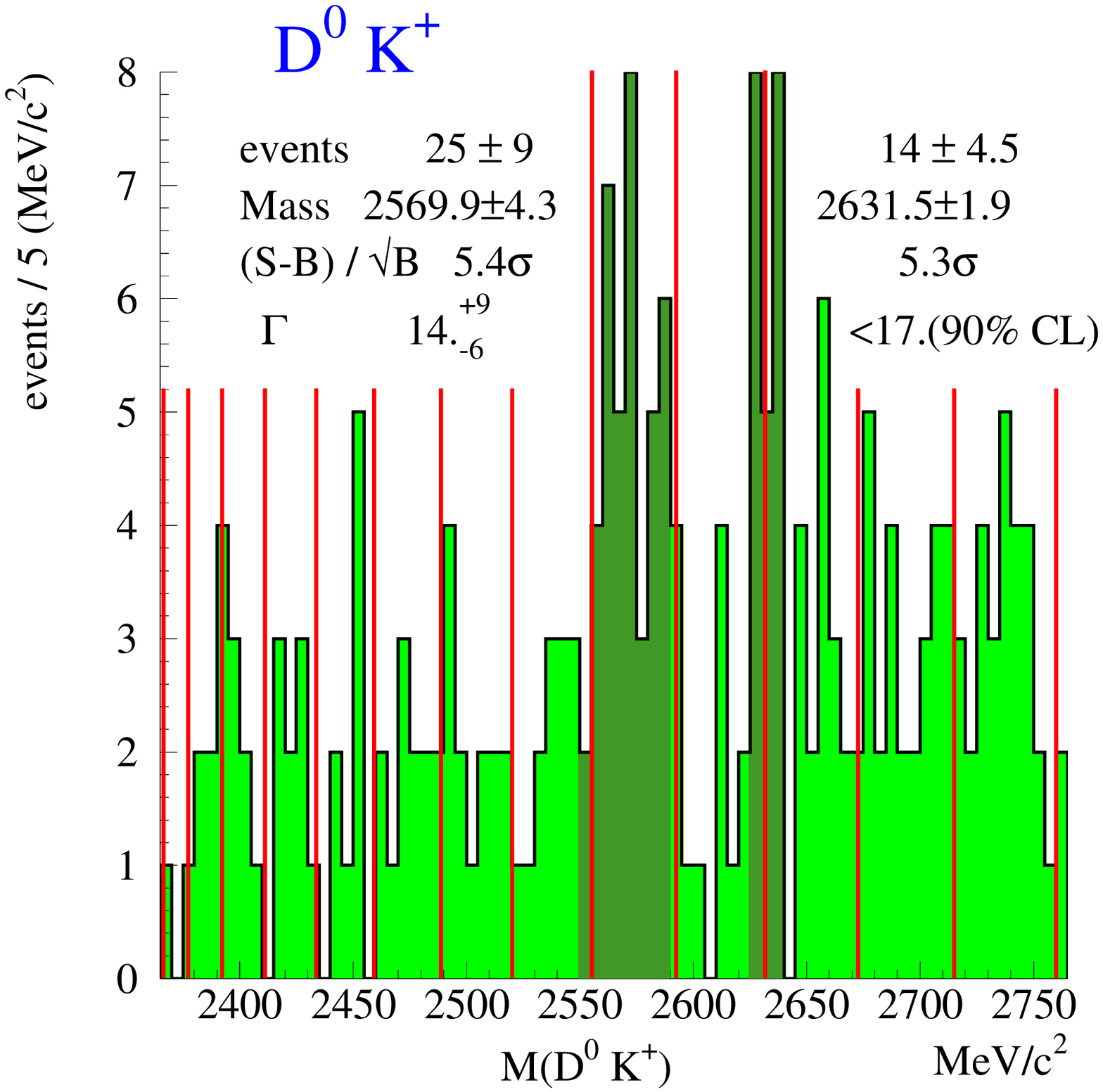}\label{fig3}
\end{center}

\caption{The $D^0K^+$ invariant mass spectrum presented in Ref.
\cite{9}. The vertical (spectral) lines correspond to KK tower for
$D^0K^+$ system; see Table 1.}
\end{figure}
\newpage

\begin{figure}[htb]
\begin{center}
\includegraphics[width=\textwidth]{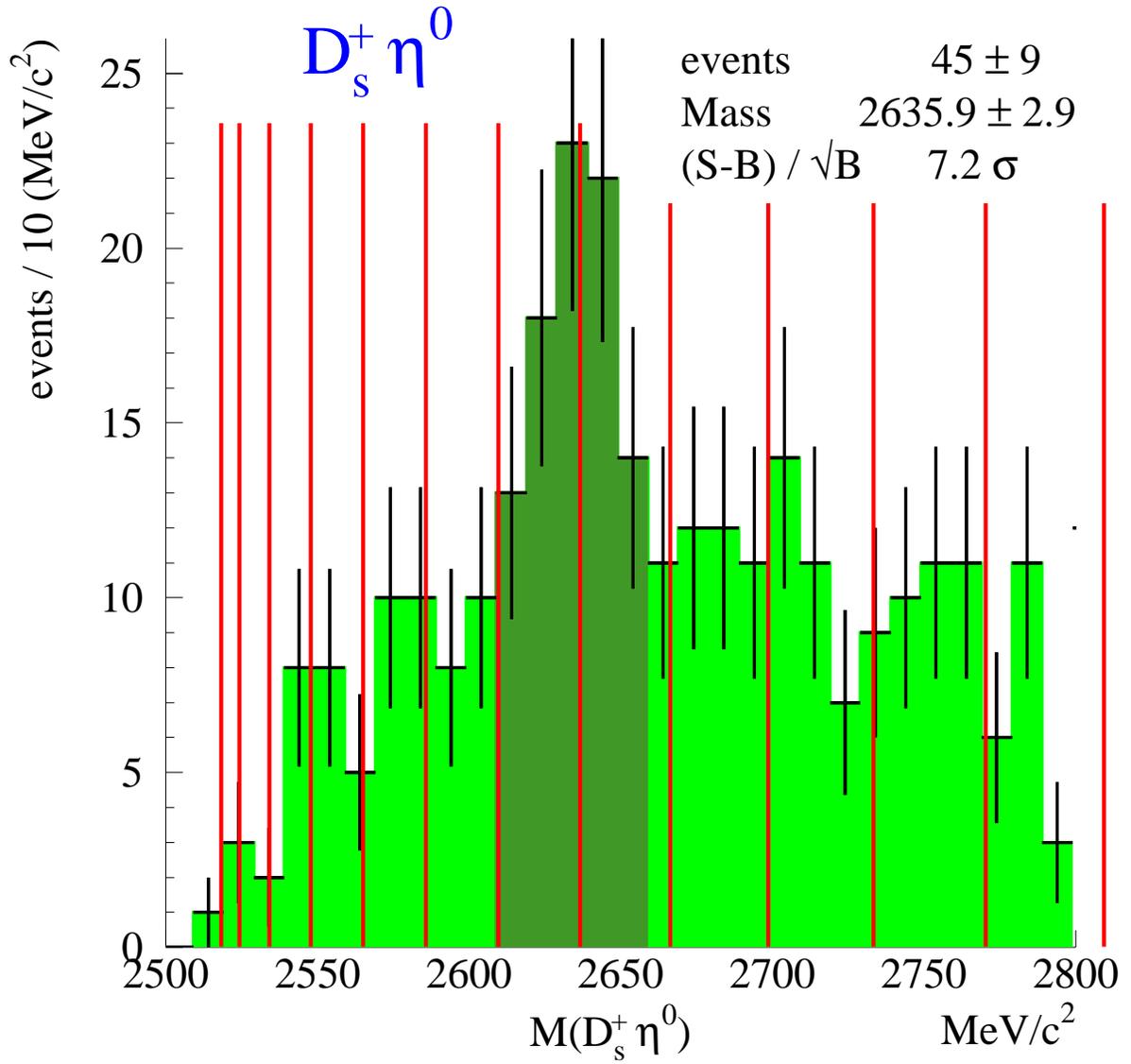}\label{fig4}
\caption{The $D_s^+\eta$ invariant mass spectrum presented in Ref.
\cite{9}. The vertical (spectral) lines correspond to KK tower for
$D_s^+\eta$ system given by Table 2.}
\end{center}
\end{figure}

\newpage

\begin{figure}[htb]
\begin{center}
\includegraphics[width=\textwidth]{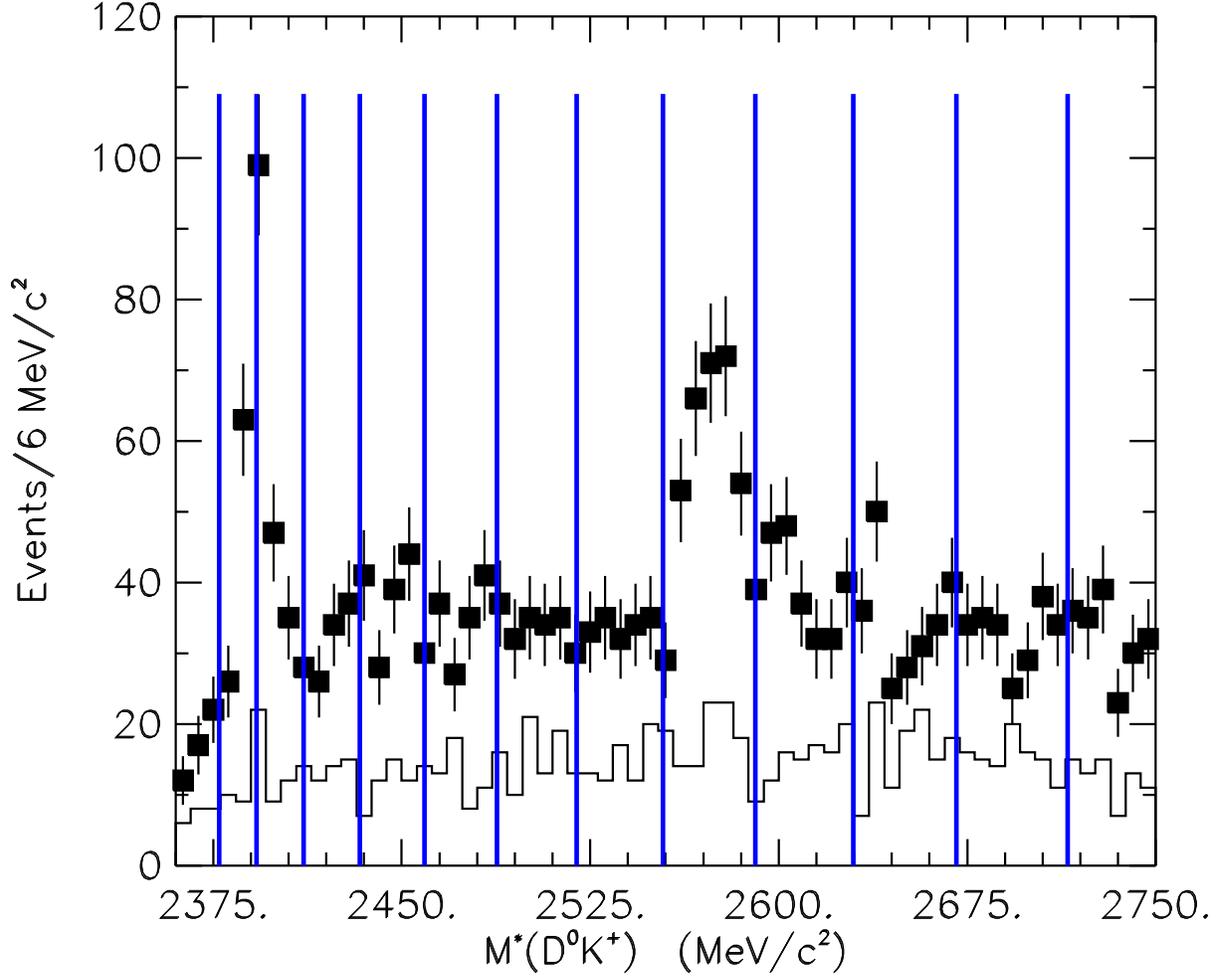}\label{fig5}
\caption{``Corrected" invariant mass of $(K^-\pi^+[\pi^0])K^+$
combinations presented in Ref. \cite{18}. Data points are for
$K^-\pi^+[\pi^0]$ combinations in the $D^0$ signal region; the
histogram shows invariant mass of $(K^-\pi^+[\pi^0])K^+$ combinations
where the $K^-\pi^+[\pi^0]$ combinations were chosen in $D^0$
sidebands. The vertical (spectral) lines correspond to KK tower for
$D^0K^+$ system given by Table 1.}
\end{center}
\end{figure}

\end{document}